\documentclass[aps,pra,twocolumn,,superscriptaddress,floatfix]{revtex4-2}
\usepackage[svgnames]{xcolor}
\definecolor{darkred}{rgb}{0.6, 0, 0}
\definecolor{darkgreen}{rgb}{0, 0.5, 0}
\usepackage[colorlinks=true,urlcolor = orange,linkcolor = darkgreen, citecolor = darkred, bookmarks=false]{hyperref}
\usepackage{amsfonts,amsmath,amssymb,amsthm}
\usepackage{graphicx}
\usepackage{dsfont}
\usepackage[normalem]{ulem}
\usepackage{enumerate}
\usepackage{natbib}
\usepackage{diagbox}

\usepackage{tcolorbox}
\usepackage{tabularx}
\usepackage{array}
\usepackage{colortbl}
\tcbuselibrary{skins}

\swapnumbers

\usepackage{tcolorbox}
\usepackage{tabularx}
\usepackage{array}
\usepackage{colortbl}
\tcbuselibrary{skins}
\newcolumntype{C}{>{\centering\arraybackslash}X}
\newcolumntype{Y}{>{\raggedleft\arraybackslash}X}

\tcbset{tab/.style={enhanced,fonttitle=\bfseries,fontupper=\small\sffamily,
		colback=green!5!white,colframe=teal,colbacktitle=red!40!white,coltitle=black,center title}}

\def\ii{{\rm i}}
\newcommand{\dd}{{\rm d}}

\usepackage{tikz}
\usepackage{pgfplots}
\usetikzlibrary{arrows,topaths,shapes.geometric,shapes.misc,patterns,positioning,shadows,decorations.markings,
decorations.pathreplacing,calc, trees, positioning, arrows, shapes, shapes.multipart, shadows, matrix, decorations.pathreplacing, decorations.pathmorphing}
\tikzset{->-/.style={decoration={
  markings,
  mark=at position #1 with {\arrow[scale=1.5]{latex}}},postaction={decorate}}}

\usepackage{pgfplots}
\pgfplotsset{compat=1.9}

\begin{document}

\title{
Dynamical Criticality of Magnetization Transfer in Integrable Spin Chains
}

\author{\v{Z}iga Krajnik}
\affiliation{Faculty for Mathematics and Physics,
University of Ljubljana, Jadranska ulica 19, 1000 Ljubljana, Slovenia}
\affiliation{CQP, Department of Physics, NYU, 726 Broadway, New York, NY 10003, United States}

\author{Johannes Schmidt}
\affiliation{Bonacci GmbH, Robert-Koch-Str. 8, 50937 Cologne, Germany}

\author{Enej Ilievski}
\affiliation{Faculty for Mathematics and Physics,
University of Ljubljana, Jadranska ulica 19, 1000 Ljubljana, Slovenia}

\author{Toma\v{z} Prosen}
\affiliation{Faculty for Mathematics and Physics,
University of Ljubljana, Jadranska ulica 19, 1000 Ljubljana, Slovenia}

\date{\today}

%%%%%%%%%%%%%%%%%%%%%%%%%%%%%%%%%%%%%%%%%
%%%%%%%%%%%%%%%%%%%%%%%%%%%%%%%%%%%%%%%%%
\begin{abstract}
Recent studies have found that fluctuations of magnetization transfer in integrable spin chains violate the central limit property. 
Here we revisit the problem of anomalous counting statistics in the Landau--Lifshitz field theory by specializing to two distinct anomalous regimes featuring a dynamical critical point. By performing optimized numerical simulations using an integrable space-time discretization we extract the algebraic growth exponents of time-dependent cumulants which attain their threshold values. The distinctly non-Gaussian statistics of magnetization transfer
in the easy-axis regime is found to converge towards the universal distribution of charged single-file systems.
At the isotropic point we infer a weakly non-Gaussian distribution, corroborating the view that superdiffusive spin transport in integrable spin chains does not belong to any known dynamical universality class.
\end{abstract}

\pacs{02.30.Ik,05.70.Ln,75.10.Jm}

\maketitle

{\bf Introduction}---Complete characterization of universal equilibrium phenomena is one of the crowning achievements  of
statistical physics \cite{Huang_book,Livi_book}. On the other hand, the rich diversity of dynamical and non-equilibrium phenomena is much harder to describe within a common general mathematical framework. Nevertheless dynamical universality has been established in certain domains such as for example noisy classical systems describing interface growth \cite{EW1982,KPZ,Quastel_2015,Takeuchi18} or, more generally, in the context of mode-coupling theory of nonlinear fluctuating hydrodynamics (NLFHD) \cite{Spohn_1991,herbert,Kulkarni2015,Popkov15}.
%A general and precise mathematical foundation however appears to be still lacking.

The study of exotic dynamical properties has been at the forefront of theoretical \cite{RevModPhys.93.025003,superdiffusion_review,GV_review} and experiment \cite{Scherg2021,Scheie21,Jepsen22,Wei2022,rosenberg2023dynamics} research in recent years. Anomalous dynamical behavior is commonly associated with a lack of ergodicity. Most prominent examples include singular diffusion constants \cite{Ljubotina19,KP20,MatrixModels,DupontMoore19,Weiner20,superdiffusion_review},
anomalous transport in kinetically constrained models \cite{PhysRevLett.127.230602} and multipole conserving systems \cite{PhysRevLett.125.245303,PhysRevE.103.022142,Knap} and Hilbert-space fragmentation \cite{PhysRevB.101.174204,PhysRevB.101.125126,Moudgalya_2022,Brighi,Motrunich_2022,PhysRevE.104.044106,Knap,GamayunBalazs}. The focus of attention has recently shifted to the study of full counting statistics and anomalous fluctuations \cite{Krajnik2022a,Krajnik2022,SarangKhemani,Krajnik2022SF,Kormos}.
For instance, classical fragmentation occurs in so-called charged single-file systems \cite{Krajnik2022SF} (including the solvable {charged hardcore lattice gas} \cite{Medenjak17} and the semi-classical low-energy regime of the sine-Gordon model \cite{Altshuler_2006,Kormos}), displaying a host of unorthodox dynamical properties such as, most prominently, a universal non-Gaussian {typical distribution of net charge transfer}.

Regarding anomalous spin transport in integrable spin chains, there are currently two elusive problems of fundamental significance
that remain unresolved. The first one concerns a first-principle microscopic justification of the Kardar-Parisi-Zhang (KPZ) scaling function \cite{Prahofer_2004} found in integrable spin chains with isotropic interactions \cite{MatrixModels,superuniversality,PhysRevLett.129.230602,superdiffusion_review}, {recently also observed experimentally~\cite{Scheie21,Wei2022}}. The second problem concerns anomalous fluctuations associated with conserved $U(1)$ charges in such models \cite{Krajnik2022a,SarangKhemani,DeNardisKPZ}, leading to a breakdown of the central limit property \cite{Krajnik2022a}. Lacking analytical tools suitable for tackling these problems, pushing the limits of numerical simulations is imperative to access the hydrodynamic regime. Classical systems are particularly suitable for this task, sidestepping the issue of rapid entanglement growth affecting their quantum counterparts.

In this work, we present a large-scale numerical study of an anisotropic classical integrable spin chain by computing the full counting statistics of cumulative spin current in thermal \emph{equilibrium}. Building upon our previous work \cite{Krajnik2022a}, an optimized numerical implementation enables us to reach longer simulation times, improving upon Refs.~\cite{Krajnik2022a,SarangKhemani}
by three orders of magnitude. Specializing to both critical regimes, we compute the time-dependent probability distribution of the cumulative spin current and extract the scaling exponents quantifying the temporal growth of cumulants.
The main findings of our study are:
(i) in the easy-axis regime, the time-dependent typical distribution slowly converges towards the universal non-Gaussian distribution characteristic of charged single-file systems \cite{Krajnik2022SF};
(ii) at the isotropic point, we infer a weakly non-Gaussian distribution and quantify small but systematic deviations from Gaussianity.

Regarding (ii) our data does not comply with a `quasi-Gaussian' distribution recently predicted in Ref.~\cite{DeNardisKPZ} within the domain of quantum spin chains. There remains the possibility that quantum fluctuations could play a pivotal role, as already alluded to in \cite{DeNardisKPZ}. However, anticipating that the `quantum-classical correspondence' \cite{Gamayun_domain,10.21468/SciPostPhys.7.2.025,Misguich19,PhysRevLett.125.070601,superdiffusion_review} persists at the level of fluctuations, a plausible consequence of our findings, including simulations of higher-rank models, is that superdiffusion in integrable isotropic chains does not admit a universal description in terms of an effective theory of nonlinear hydrodynamics.

{\bf Model}---We consider the anisotropic Landau-Lifshitz magnet in thermal equilibrium at high temperature.
The time evolution of a classical spin field ${\bf S}\equiv (S^{1},S^{2},S^{3})^{\rm T} \in \mathcal{S}^2$ is governed by
\begin{equation}
	\partial_{t}{\bf S} = {\bf S}\times \partial^{2}_{x}{\bf S} + {\bf S}\times {\mathrm J}\,{\bf S},
	\label{eqn:LL}
\end{equation}
with anisotropy tensor $\mathrm{J}={\rm diag}(0,0,\delta)$ parametrized by $\delta \in \mathbb{R}$.
Equation \eqref{eqn:LL} is a prime example of a completely 

\begin{widetext}
	
	\begin{figure}[t]
		\centering
		\includegraphics[width=\columnwidth]{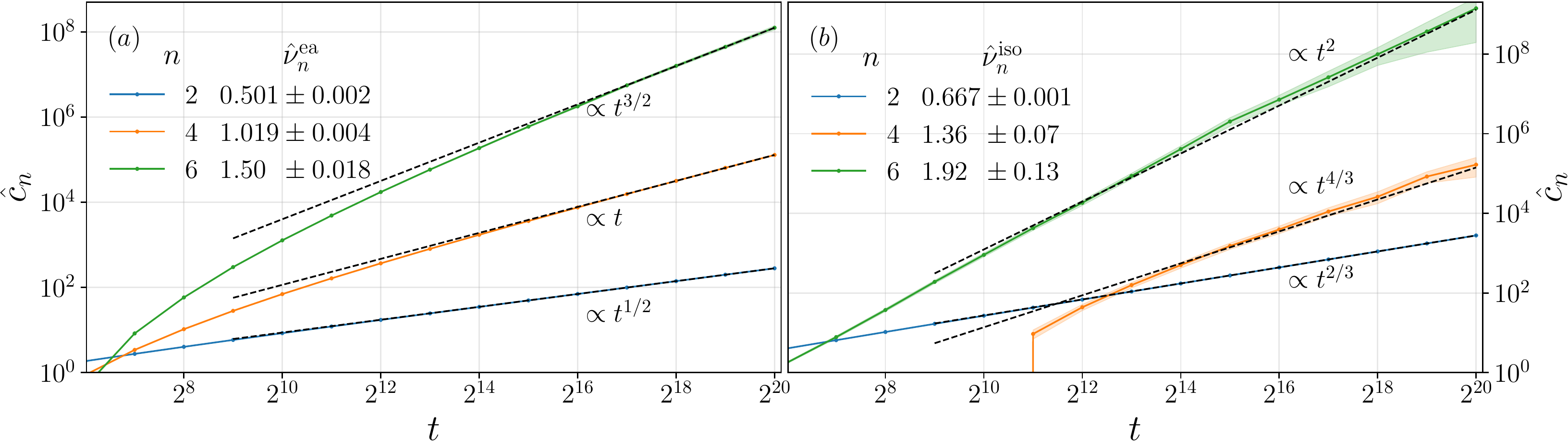}
		\caption{Temporal growth of cumulant estimates $\hat c_n(t)$ for $n\in \{2,4,6\}$ (colored dots)
		with three standard deviation neighborhoods (shaded regions): (a) easy-axis regime ($\varrho = 1$), (b) isotropic point ($\varrho=0$).	
		Dashed black lines show algebraic scalings \eqref{c_scaling} with fitted exponents
		(a) $\nu_n^{\rm ea}$, given by Eq.~\eqref{ea_nu}, and (b) $\nu_n^{\rm iso}$, given by Eq.~\eqref{iso_nu}.
		Finite-sample exponents $\hat \nu_{n}$ are estimated from finite-time data in the time interval $t \in [2^{16}, 2^{20}]$.
		Simulation parameters: $\tau = 1$, $L = 2^{21}$, $N = 5\times10^3$ (see Ref.~\cite{LLMM}).}
		\label{fig:figure1}
	\end{figure}
\end{widetext}
integrable PDE~\cite{Takhtajan77,Borovik78,Faddeev_book,BikbaevBobenkoIts2014}. It possesses infinitely many local conserved quantities, including 
the third component of total spin $Q=\int \dd x\,S^{3}(x,t)$
with the spin density $S^3$ obeying the continuity equation $\partial_{t}S^{3}(x,t)+\partial_{x}j(x,t)=0$,  where $j(x,t)$ denotes the spin-current density. By tuning $\delta$, one can access three distinct dynamical regimes: ({\tt ea}) the easy-axis regime ($\delta > 0$), ({\tt iso}) the isotropic point ($\delta=0$) and ({\tt ep}) the easy-plane regime ($\delta < 0$).

{\bf Anomalous statistics of magnetization transfer}---Our study mainly concerns the time-dependent distribution $\mathcal{P}(J|t)$ of
the cumulative  current ${J}(t) = \int^{t}_{0}\dd t'\,j(0,t')$
passing through the origin in a finite time interval of length $t$. Defining the moment generating function (MGF)
$G(\lambda|t) \equiv \langle e^{\lambda J(t)} \rangle = \int \dd J \mathcal{P}(J|t) e^{\lambda J}$, where
the average $\langle \bullet \rangle$ is computed in a maximum entropy/infinite temperature ensemble,
we characterize $\mathcal{P}(J|t)$ by its cumulants, $c_n(t) \equiv \frac{\dd^n}{\dd \lambda^n} \log G(\lambda|t) \big|_{\lambda=0}$.
We assume that $c_{n}(t)$ grow asymptotically with time as
\begin{equation}\label{c_scaling}
c_n(t) \asymp c_{n} t^{\nu_n},
\end{equation}
with algebraic growth exponents $\nu_{n}$. Moreover, time-reversal symmetry in equilibrium ensembles implies detailed balance, reflected in the symmetry $\mathcal{P}(J|t)=\mathcal{P}(-J|t)$. Accordingly, all odd cumulants vanish.

The growth of variance (second cumulant) $c_{2}(t)$ determines the \emph{typical} timescale of magnetization transfer with exponent $\nu_2 = 1/z$, given by the \emph{dynamical exponent} $z$, governing the hydrodynamic relaxation of
the density two-point function. By accordingly rescaling the cumulative current, $\mathcal{J}(t) \equiv t^{-1/2z}J(t)$,
the $t\to \infty$ limit of the rescaled distribution $\mathcal{P}_{1/2z}(\mathcal{J}|t) \equiv t^{1/2z} \mathcal{P}(\mathcal{J}|t)$
yields the \emph{typical distribution} $\mathcal{P}_{\rm typ}(j) \equiv \lim_{t \to \infty} \mathcal{P}_{1/2z}(\mathcal{J}=j|t)$.
The second and higher cumulants of the finite-time typical distribution $\kappa_n(t) \equiv \langle \left[ \mathcal{J} (t)\right]^n \rangle^c$ are directly related to $c_{n}(t)$ by a simple rescaling, $\kappa_{n}(t) = t^{-n/2z} c_n(t)$.

%{\bf Dynamical criticality}---
Time-dependent cumulants $c_{n}(t)$ generically exhibit linear asymptotic growth, $c_{n}(t)\sim t$ for
all even $n$. In such a \emph{regular} scenario, the central limit property follows from formal analytic properties of the MGF (see Refs.~\cite{Krajnik2022a,Krajnik2022} for a detailed discussion), implying that typical fluctuations are normally distributed,
$\mathcal{P}_{\rm typ} = \mathcal{N}(0, \kappa_2)$. By contrast, in a \emph{dynamically critical} scenario,
$G(\lambda|t)$ experiences an \emph{equilibrium} dynamical phase transition at the \emph{critical} counting field $\lambda_{c}=0$, causing in effect a superlinear growth of higher cumulants, namely $\lim_{t\to \infty}c_{m}(t)/t \to \infty$ for some $m>2$. Such dynamical criticality can be quantified by the algebraic growth exponents $\nu_{n}$, see Eq.~\eqref{c_scaling}. If the exponents take the threshold values $\nu^{\rm thr}_{n}=n/2z$, $\mathcal{P}_{\rm typ}$ will be non-Gaussian (exactly solvable examples are discussed in Refs.~\cite{Krajnik2022,Krajnik2022SF}).
 
{\bf Methods}---In the present work, we find clear signatures of dynamical criticality in the easy-axis ({\tt ea}) and isotropic ({\tt iso}) regimes of the anisotropic Landau-Lifshitz theory \eqref{eqn:LL}, thereby corroborating the earlier results of Ref.~\cite{Krajnik2022a}.
In addition, we here numerically extract the growth exponents $\nu_n$ and quantify the emergent
typical distributions $\mathcal{P}_{\rm typ}$ in both critical regimes ($\delta \geq 0$).

To enhance efficiency and to avoid potential artifacts stemming from naive discretizations of Eq.~\eqref{eqn:LL}, we perform our simulations using a two-parameter integrable symplectic discretization of Eq.~\eqref{eqn:LL} developed in Ref.~\cite{LLMM}, depending on anisotropy parameter $\varrho$ and time-step parameter $\tau$ (definitions and further details can be found in \cite{LLMM}, and the Supplemental Material of Ref.~\cite{Krajnik2022a}). Simulations were performed on periodic systems of length $L = 2^{21} \geq 2t_{\rm max}$ with maximal time $t_{\rm max}=2^{20}$, to exclude finite-size effects. We subsequently use hatted symbols $\hat \bullet$ to denote finite-sample estimates of ensemble-averaged quantities.
The time-dependent moments $m_n(t)\equiv(\dd/\dd \lambda)^{n} G(\lambda|t) |_{\lambda=0}$ of the discrete cumulative current $J_\ell^t$ were
estimated as $\hat m_n(t) = (L\,N)^{-1} \sum_{s=1}^{N} \sum_{\ell=1}^{L}  \left(J_\ell^t[s]\right)^n$,
using $[s]$ to denote the $s$-th trajectory taken from an ensemble of $N$ samples, such that
$\lim_{N \to \infty} \hat m_n(t) = m_n(t)$, while the inner sum exploits translational invariance to
improve sampling statistics. The estimated cumulants $\hat c_n$ are computed directly from $\hat m_n$ using Fa\`{a} di Bruno's formula.
To quantify the proximity between a continuous distribution $\mathcal{P}$ and a target distribution $\mathcal{Q}$ we utilize the Kullback-Leibler (KL) divergence $D_{\rm KL}(\mathcal{P}||\mathcal{Q}) \equiv \int_{-\infty}^{\infty} \dd x\, \mathcal{P}(x)\log \left[{\mathcal{P}(x)}/{\mathcal{Q}(x)}\right]$. The unknown estimated widths of the asymptotic target distributions, denoted by $\hat{\sigma}$,
are extracted by means of a non-linear least squares fit to the finite-time distributions at $t_{\rm max}$.
  
\begin{figure}[t]
	\centering
	\includegraphics[width=\columnwidth]{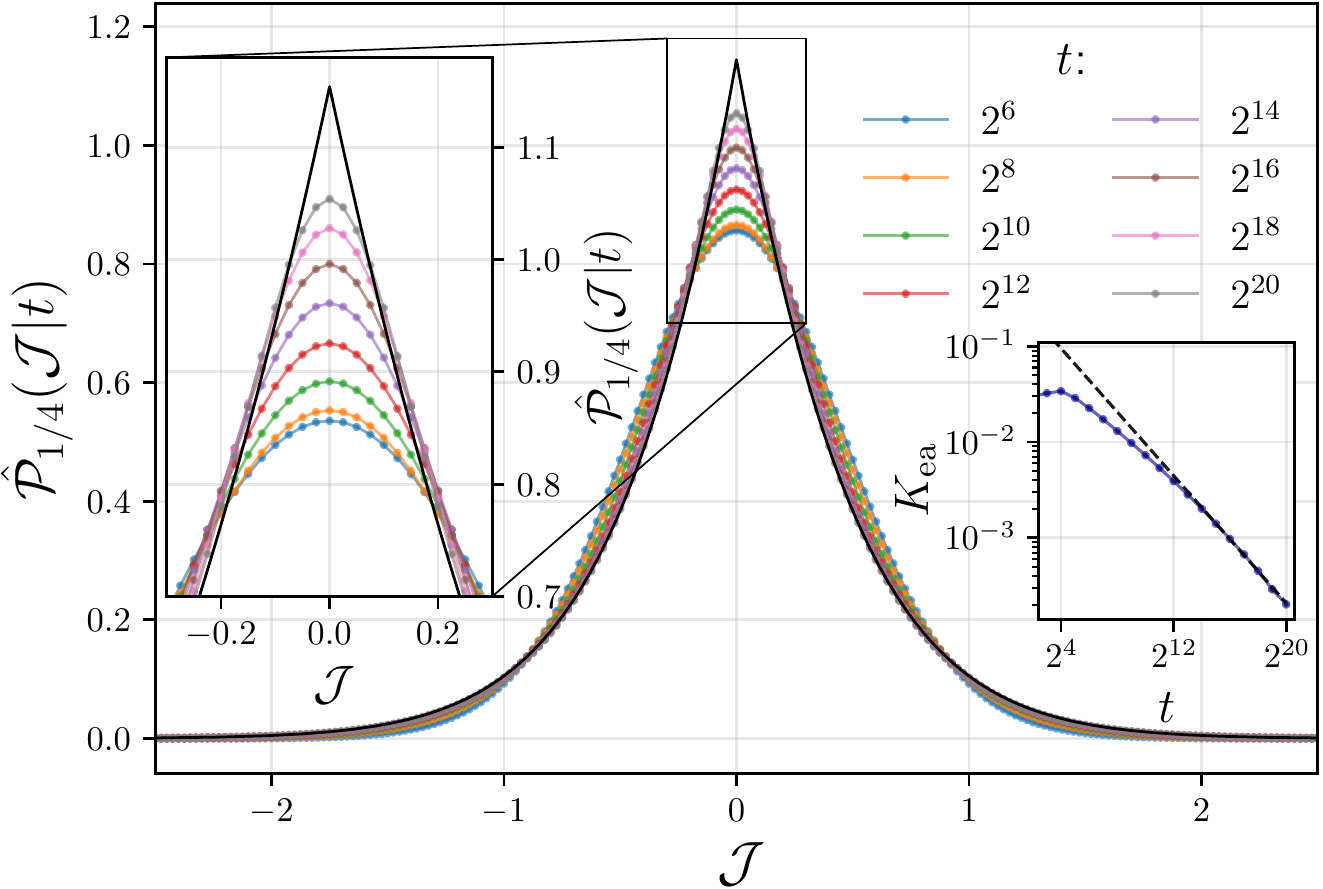}
	\caption{Convergence of the estimated time-dependent distributions $\hat{\mathcal{P}}_{1/4}(\mathcal{J}|t)$ (colored points)
	in the easy-axis regime ($\varrho=1$) towards the conjectured target distribution $\mathcal{M}_{\hat \sigma_{\rm ea}}$ \eqref{sf_dist}, with $\hat \sigma_{\rm ea}\approx0.3595$ (solid black curve), with a zoom-in near the origin (left inset).
	(right inset) Relaxation of KL divergence $K_{\rm ea}(t)$ (blue), with an estimated asymptotic decay $\sim t^{-0.55}$
	(dashed black line) fitted for $t\geq 2^{15}$. Simulation parameters: $\tau = 1$, $L = 2^{21}$, $N=4\times 10^{3}$.}
	\label{fig:figure2}
\end{figure}

{\bf Easy-axis regime}--- In the easy-axis regime ($\delta>0$), we confirm the anticipated critical behavior of $c_{n}(t)$ across four orders of magnitude in time.  Temporal growth of the few lowest even cumulants $c_{n}(t)$ is shown in Fig.~\ref{fig:figure1}a, from where we deduce the growth exponents
\begin{equation}\label{ea_nu}
 \nu_{2n}^{\rm ea} = n/2.
\end{equation}
This readily implies non-zero cumulants of the typical distribution,
${\kappa}_{n} = \lim_{t \to \infty} \lim_{N \to \infty} \hat \kappa_{2n}^{\rm ea}(t) \neq 0$.
As shown in Fig.~\ref{fig:figure2},
the finite-time typical distributions $\hat{\mathcal{P}}_{1/4}(\mathcal{J}|t)$ are discernibly non-Gaussian, converging at late times towards
the M-Wright distribution $\mathcal{M}_{\sigma}(j) \equiv \sigma^{-1/2}M_{1/4}(2|j|/\sigma^{1/2})$ \cite{Mainardi_2020},
given by the following explicit integral representation \cite{Krajnik2022a,Krajnik2022SF}
\begin{equation}\label{sf_dist}
	\mathcal{M}_{\sigma}(j) = \int_{-\infty}^{\infty} \frac{\dd u }{2 \pi \sigma |u|^{1/2}}  \exp \left[ -\frac{u^2}{2\sigma^2} - \frac{j^2}{2|u|}\right].
\end{equation}
Convergence of $\hat{\mathcal{P}}_{1/4}$ towards $\mathcal{M}_{\hat{\sigma}_{\rm ea}}$ near the origin
%(with $\hat{\sigma}_{\rm ea}$ estimated at $t_{\rm max}$) 
is displayed in Fig.~\ref{fig:figure2} (left inset). The KL divergence $K_{\rm ea}(t) \equiv D_{\rm KL}(\hat{\mathcal{P}}_{1/4}|| \mathcal{M}_{\hat \sigma_{\rm ea}})$ decays approximately as $K_{\rm ea}(t)\sim t^{-0.55}$ with $K_{\rm ea}(t_{\rm max})\approx 2.0\times 10^{-4}$,
see Fig.~\ref{fig:figure2} (right inset).

\begin{figure}[t]
	\centering
	\includegraphics[width=\columnwidth]{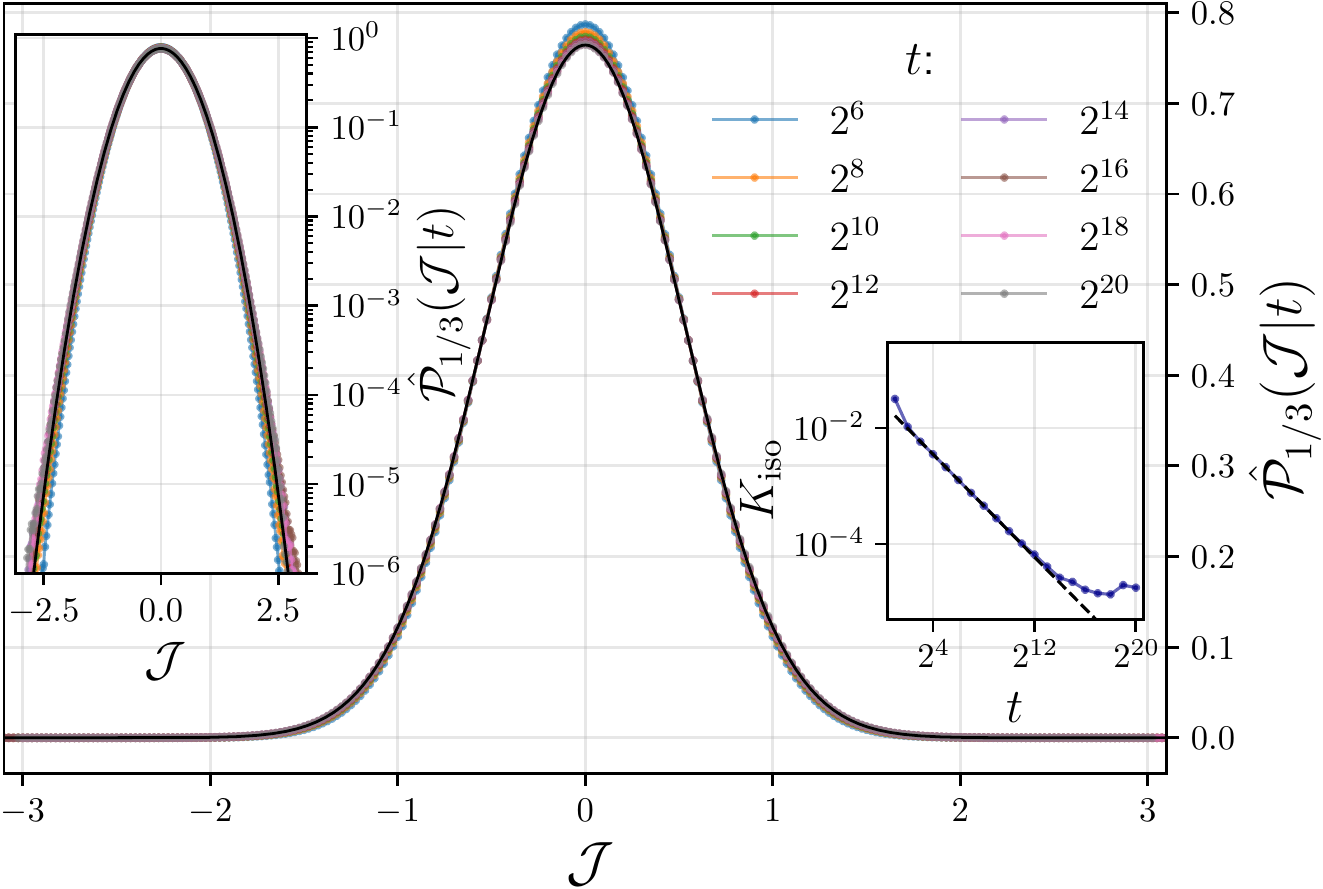}
	\caption{Convergence of the estimated time-dependent distributions $\hat{\mathcal{P}}_{1/3}(\mathcal{J}|t)$  (colored points) at the isotropic point ($\varrho=0$), compared against the Gaussian distribution $\mathcal{N}_{\sigma_{\rm iso}}$ with $\hat \sigma_{\rm iso} \approx 0.522$ (solid black curve) in logarithmic scale (left inset).
	(right inset) Relaxation of KL divergence $K_{\rm iso}(t)$ (blue), with approximate algebraic decay $\sim t^{-0.74}$ (black dashed line),
	fitted in the window $t\in [2^{3},2^{14}]$. Simulation parameters: $\tau = 1$, $L = 2^{21}$, $N=4\times 10^{3}$.}
	\label{fig:figure3}
\end{figure}

The probability distribution \eqref{sf_dist} of charge fluctuations in (unbiased) equilibrium states has been recently established in \cite{Krajnik2022SF} as one of the defining universal properties of classical charged single-file systems.
While our data empirically demonstrates convergence towards $\mathcal{M}_{\sigma}(j)$, it is important to emphasize
that the single-file constraint is not (at least manifestly) present in the easy-axis regime of our model.
In other words, the emergence of $\mathcal{M}_{\sigma}(j)$ is not a direct corollary of a kinetic constraint.
Nonetheless, Ref.~\cite{SarangKhemani} explains how \eqref{sf_dist} arises from a simple phenomenological hydrodynamic picture based on elastic scattering of magnons off immobile domain walls describing the large-anisotropy regime of the gapped Heisenberg spin chain, suggesting that \eqref{sf_dist} is not exclusive to single-file systems but also allows for a finite transmission rate. While this viewpoint is also alluded to in \cite{Kormos}, a systematic or rigorous derivation is currently still lacking.

{\bf Isotropic point}---As shown in Fig.~\ref{fig:figure1}b, dynamical criticality persists at the isotropic point ($\delta=0$).
Once again the numerically extracted first few growth exponents match the threshold values
\begin{equation}\label{iso_nu}
	 \nu_{2n}^{\rm iso} = 2n/3,
\end{equation}
implying that the typical distribution acquires non-zero cumulants ${\kappa}^{\rm iso}_{2n}=\lim_{t \to \infty} \lim_{N \to \infty} \hat \kappa_{2n}^{\rm iso}(t) \neq 0$, signaling a breakdown of the central limit property.

We next quantify how much $\mathcal{P}^{\rm iso}_{\rm typ}$ deviates from Gaussianity.
A direct quantitative comparison between the estimated time-dependent typical distribution $\hat{\mathcal{P}}_{1/3}$ and
a normal distribution $\mathcal{N}_{\hat{\sigma}_{\rm iso}}\equiv \mathcal{N}(0, \hat \sigma_{\rm iso}^2)$ with the estimated variance $\hat \sigma_{\rm iso}\approx 0.522$, shown in Fig.~\ref{fig:figure3}, is rather nuanced:  The distance between distributions clearly decreases with time across six orders of magnitude, see Fig.~\ref{fig:figure3} (left inset), and the KL divergence $K_{\rm iso}(t) \equiv D_{\rm KL}(\hat{\mathcal{P}}_{1/3}||\mathcal{N}_{\hat \sigma_{\rm iso}})$ decays approximately as $K_{\rm iso}(t) \simeq t^{-0.74}$ at large intermediate times before crossing over to a plateau around $t\approx 2^{17}$ ($K_{\rm iso}(t_{\rm max}) \approx 1.7\times10^{-5}$, see right inset in Fig.~\ref{fig:figure3}).
It is unclear if such behavior persists for times beyond $t_{\rm max}$.

\begin{figure}[t]
	\centering
	\includegraphics[width=\columnwidth]{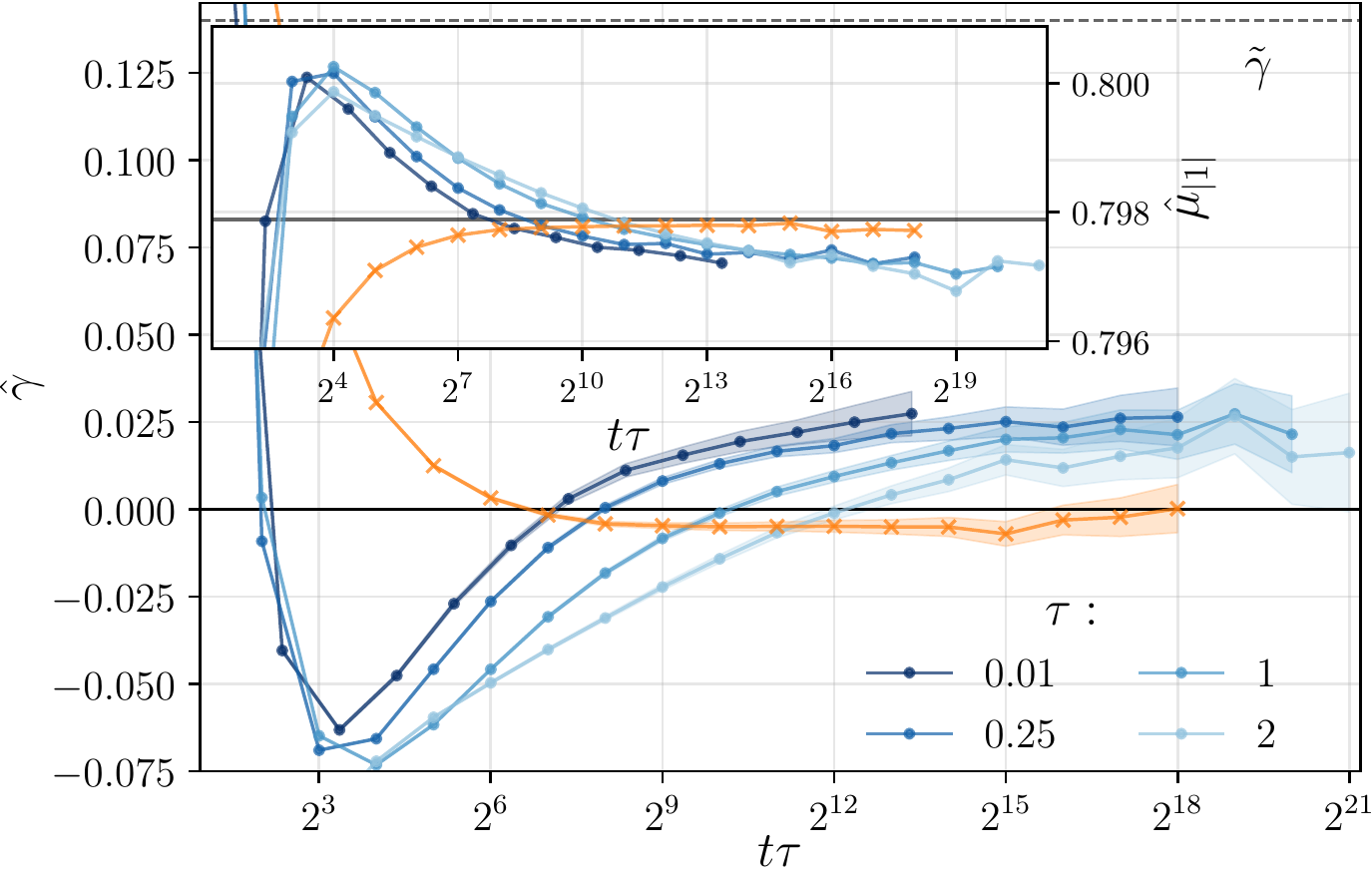}
	\caption{Numerically estimated excess kurtosis $\hat \gamma(t\tau)$ and standardized first absolute moment $\hat \mu_{|1|}(t\tau)$ (inset) at the isotropic point ($\varrho=0$)
	with three standard deviation neighborhoods (shaded regions), shown for different $\tau$ (blue points) and for $\mathbb{CP}^{2}$ (see main text) with $\tau=1$ (orange crosses). Solid black lines indicate Gaussian values $\gamma^{\mathcal{N}} = 0$ and $\mu_{|1|}^{\mathcal{N}} = \sqrt{2/\pi}$ (inset), while
	dashed black line marks $\tilde \gamma \approx 0.14$ of Ref.~\cite{DeNardisKPZ}.
	Simulation parameters: $\tau \in \{0.01, 0.25, 1, 2\}$, $L = 2^{21}$, $N\in [10^{3},5\cdot 10^{3}]$,
	with more samples for larger $\tau$.}
	\label{fig:figure4}
\end{figure}

Unlike in the easy-axis regime, the difference primarily builds up in the tails.
To discriminate between the estimated and target distribution, we compute the excess kurtosis $\hat \gamma(t) =\hat c_4(t)/\hat c_2^2(t)$ and the standardized first absolute moment $\hat \mu_{|1|}(t) = \hat m_{|1|}(t)/\hat{c}_2^{1/2}(t)$ of $\hat{\mathcal{P}}_{1/3}$.
In addition, we compare the results of our simulations with the recent prediction of Ref.~\cite{DeNardisKPZ} which reports the
(approximate) asymptotic value $\tilde \gamma \approx 0.14$.
%and $\tilde \mu_{|1|} \approx 0.788$.
In our simulations, see Fig.~\ref{fig:figure4}, we instead obtain $\hat \gamma(t_{\rm max}) \approx 0.02$ and
$\hat \mu_{|1|}(t_{\rm max}) \approx 0.7972$, see inset of Fig.~\ref{fig:figure4}.  The estimated kurtosis $\hat \gamma$ agrees with values obtained from experiments on quantum simulators \cite{rosenberg2023dynamics}.
Most glaringly, we find no decay towards the Gaussian values $\gamma^{\mathcal{N}} = 0$ and $\mu_{|1|}^{\mathcal{N}} = \sqrt{2/\pi}\approx 0.7979$.

To check whether the small value of kurtosis is universal, we also consider Noether charge fluctuations in a (generalized) $SU(N)$ Landau--Lifshitz model on the complex projective space $\mathbb{CP}^{N-1}$(specializing to $N=3$) (see \cite{MatrixModels,suppmat}). We find dynamically critical cumulants (not shown) with threshold exponents identical to those in Eq.~\eqref{iso_nu}. The corresponding standardized first absolute moment and excess kurtosis (orange crosses in Fig.~\ref{fig:figure4} with $t_{\rm max}^{(3)}=2^{18}$) are again non-zero, but distinct from those in the $\mathbb{CP}^1$ model and substantially closer to Gaussian values $\hat \mu_{|1|}^{(3)}(t_{\rm max}^{(3)}) \approx 0.7977$ and $\hat \gamma^{(3)}(t_{\rm max}^{(3)}) \approx -4\times10^{-3}$. 

{\bf Easy-plane regime}---In the easy-plane regime (with ballistic exponent $z^{\rm ep}=1$), even cumulants grow linearly with time on intermediate timescales, $\nu_{2n}^{\rm ep} = 1$.
This behavior is consistent with regularity and thus the analytical prediction
of ballistic macroscopic fluctuation theory \cite{MBHD20,DoyonMyers20,BMFT} is expected to be valid in this regime. However, a reliable extraction of scaled cumulants $s_n = \lim_{t \to \infty} t^{-1}c_n(t)$ is difficult in practice due to the required exact cancellation of $n-1$ leading orders. 

%Perhaps counterintuitively, an accurate numerical estimation of cumulants requires more computational resources in the regular scenario. To see this, we expand the finite-time moment estimates as $\hat m_n(t) \simeq \sum_{k=0}^n \hat m_{n, k}  t^{k/2z}$.
%Finite-time cumulants are then expressed as $\hat c_n(t) \simeq \sum_{k=0}^n \hat c_{n,k} t^{k/2z}$,
%where the expansion coefficients $\hat c_{n,k}$ are homogeneous (in the degree $n$ and order $k$) polynomials of coefficients $\hat m_{n,k}$. 
%We can now recognize that linear growth $\hat c_n^{\rm reg}(t) \sim \hat c_{n,1} t$ in the regular case
%requires an exact cancellation of all $n-1$ leading orders, $\lim_{N \to \infty} \hat c_{n, n\geq k>1} = 0$. In a finite sample however, \emph{all} the coefficients are invariably non-zero due to statistical fluctuations, and for large times even tiny deviations in the leading-order coefficients are rapidly amplified.
%To reliably extract $\hat c_{n,1}$, one must ensure that $\hat c_{n, k} t^{k-1} \ll \hat c_{n, 1}$ at all orders $n \geq k> 1$.
%In the critical case, the asymptotic scaling of cumulants is governed by the leading coefficients,
%$\hat c_n^{\rm crit}(t) \sim \hat c_{n,n} t^{n/2z}$. To accurately extract $\hat c_{n, n}$ it is therefore suffices
%ensuring that $\hat c_{n,k}t^{(k-n)/{2z}} \ll \hat c_{n,n}$ at all orders $n>k\geq 0$. Unlike in the regular case, the left-hand side now decays with time.

{\bf Conclusion and discussion}---In this work, we studied statistical properties of magnetization transfer in the Landau--Lifshitz field theory,
focusing on unbiased equilibrium states. Using an efficient implementation of an integrable space-time discretization,
we numerically estimated a few lowest cumulants and extracted the algebraic exponents quantifying their temporal growth.
In the easy-axis regime and at the isotropic point, the onset of dynamical criticality causes super-linear growth of higher cumulants.
In both cases, the estimated growth exponents coincide with threshold values, suggesting a violation of the central limit property. 

In the easy-axis regime, typical fluctuations are distinctly non-Gaussian, and our data convincingly demonstrates convergence towards the M-Wright distribution featured in charged single-file systems. This finding conforms with the prediction of the phenomenological model of Ref.~\cite{SarangKhemani} describing the large-anisotropy limit of the gapped Heisenberg quantum spin chain.
At the isotropic point the lowest standardized moments of the time-dependent typical distribution are found to converge close to Gaussian values, but we still detect systematic deviations that persist at late times.

%To the best of our knowledge, there is currently no generally accepted definition of dynamical universality.
%The KPZ universality class is conventionally discussed in the domain of interface growth models, referring to
%a unified coarse-grained evolution of the fluctuating height field in a particular scaling regime \cite{Corwin2012,Quastel_2015,baik2022kpz}, yielding an Airy process \cite{TracyWidom94,PrahoferSpohn02}.
%Such a limit is unfortunately difficult to perform, and thus one instead most often focuses on simple dynamical observables such as local fluctuations of charge or current densities.

A commonly used classification of dynamical universality within the framework of NLFHD \cite{Popkov15,Popkov16,Schutz18,PhysRevE.100.052111} is based on the asymptotic form of dynamical two-point functions (dynamical structure factors), characterized by an algebraic decay exponent and stationary scaling profiles.
Such a classification is however not exhaustive, as different processes may only be distinguishable at the level of higher-order dynamical correlation functions such as e.g. the full counting statistics of charge transfer studied in this work. A subclass of \emph{ballistic} charged single-file systems provides an illustrative example \cite{Krajnik2022}: while magnetization transport in unbiased equilibrium states yields diffusive (i.e. Gaussian) scaling profiles \cite{Klobas2018}, statistics of magnetization transfer is anomalous and
described by the distribution \eqref{sf_dist}, ruling out normal diffusion.

A similar pitfall arises when classifying superdiffusive transport of non-abelian charges in integrable systems \cite{Vir20,MatrixModels,superuniversality,superdiffusion_review}. There is by now ample numerical evidence \cite{Ljubotina19,DupontMoore19,KP20,MatrixModels,Das19_KPZ,Weiner20,superuniversality,PhysRevLett.129.230602} 
that the asymptotic structure factors yield the Pr\"{a}hofer-Spohn scaling function \cite{Prahofer_2004} of the KPZ universality class, referring to a unified coarse-grained description of the fluctuating height field in a scaling regime of interface growth models \cite{Corwin2012,Quastel_2015,baik2022kpz}.

 However, as  originally pointed out in Ref.~\cite{Krajnik2022a}, the KPZ equation initialized in the stationary ensemble \cite{Ferrari06,Imamura12,Corwin2012} generates inherently asymmetric fluctuations \cite{Baik2000,Baik_2001} due to broken detailed balance, contrasting with the situation in integrable spin chains.
 While hydrodynamic equations involving two coupled KPZ modes \cite{DeNardisKPZ} resolve this shortcoming, our numerical simulations reveal systematic deviations from both the Gaussian values and the two-mode quasi-Gaussian distribution.

The non-universal estimated values of kurtosis at the isotropic point plausibly suggest that either
(i) fluctuations (i.e. higher-point temporal correlations of current densities) in integrable spin chains with non-abelian symmetries
are, unlike the dynamical structure factor, dependent on the symmetry group, or (ii) that higher standardized
moments eventually relax to (presumably Gaussian) values on extremely long timescales inaccessible to current numerical simulations,
indicating non-algebraic (e.g. logarithmic) corrections to critical cumulant scaling \eqref{c_scaling}.

Our work raises several important questions. It remains to be examined whether quantum corrections alter the observed classical phenomenology and how symmetries are reflected in higher-point correlations.
 It likewise remains unclear whether integrable systems featuring infinitely many local conserved quantities permit a reduction to effective mode-coupling equations with finitely many modes. Another important question to be explored concerns the structure of large-deviation rate functions in both dynamically critical regimes and whether first and second order dynamical phase transitions found in charged single-file systems \cite{Krajnik2022SF} manifest themselves in the easy-axis or isotropic regimes away from equilibrium.

We would like to close by highlighting the fact that many problems concerning the counting statistics of charge transfer in \emph{quantum}
systems are now finally within reach of contemporary experimental techniques, as recently exemplified in \cite{rosenberg2023dynamics}. We are hopeful that quantum simulators can provide valuable insights.
 
\paragraph*{\bf Acknowledgments}
We are indebted to J. De Nardis for valuable correspondence and for suggesting the computation of the standardized first absolute moment.
We thank S. Gopalakrishnan, R. Vasseur, V. Popkov and G. Sch\"{u}tz for insightful discussions and V. Pasquier for collaboration on related projects.
This work has been supported by the Slovenian Research Agency (ARRS) under the Program P1-0402 and grants N1-0233 and N1-0219. \v{Z}K acknowledges support of the Milan Lenar\v{c}i\v{c} Foundation and the Simons Foundation via a Simons Junior Fellowship grant 1141511. EI is supported by grant N1-0243 of ARRS.

%%%%%%%%%%%%%%%%%%%%%%%%%%%%%%%%%%%%%%%%%
%%%%%%%%%%%%%%%%%%%%%%%%%%%%%%%%%%%%%%%%%

\bibliography{universality_classes}

\clearpage

\onecolumngrid
\begin{center}
	\textbf{{\large Supplemental Material for \\ ``Critical Fluctuations of Magnetization Transfer in Integrable Spin Chains''}}
\end{center}

\section{Generalized isotropic Landau--Lifshitz theories}
Generalized isotropic  ($SU(N)$) Landau--Lifshitz theories describe matrix degrees of freedom ${\rm M} \in {\rm Gr}_{\mathbb{C}}(k, N)$ where
\begin{equation}
	{\rm Gr}_{\mathbb{C}}(k, N) = \left\{{\rm M} \in GL(N, \mathbb{C})|\enspace {\rm M}={\rm M}^{\dagger},\ {\rm M}^{2} = \mathds{1}_N,\ {\rm Tr}{\rm M} = N-2k\right\}.
\end{equation}
is the complex Grassmanian manifold and $k$ denotes the rank of ${\rm M}$. Grassmanian manifolds of rank one ($k=1$) are isomorphic to complex projective spaces, ${\rm Gr}_{\mathbb{C}}(1, N) \simeq \mathbb{CP}^{N-1}$.  Observables ${\rm M}^a$ on the phase-space are given by the moment map
\begin{equation}
	{\rm M}^a = {\rm Tr}(X^a {\rm M}), \quad a = 1, 2, \ldots N^2-1,
\end{equation}
where $X^a \in \mathfrak{su}(N)$ are the $SU(N)$ generators normalized as ${\rm Tr}(X^a X^b) = 2\delta_{ab}$.
The dynamics of ${\rm M}(x, t)$ is given by
\begin{equation}
	\partial_t {\rm M} = \frac{1}{2\ii}\left[{\rm M}, \partial_x^2 {\rm M} \right]. \label{M_eom}
\end{equation}
Eq. \eqref{M_eom} is an integrable PDE for all $N\in \mathbb{N}$, $k \in \mathbb{N}_{N}$. To avoid breaking integrability, we perform numerical simulations of generalized Landau--Lifshitz theories \eqref{M_eom} with integrable symplectic space-time discretization developed in \cite{MatrixModels}. The $\mathbb{CP}^2$ simulation presented in Figure 4 of the main text correspond to $N=3$ and $k=1$.

\subsection*{Noether charge fluctuations}
The $SU(N)$ symmetry of \eqref{M_eom} generates conserved Noether charges
\begin{equation}
	Q^a = \int \dd x\,  M^a(x, t),
\end{equation}
whose densities satisfy local continuity equations
\begin{equation}
	\partial_t {\rm M}^a + \partial_x j^a(x, t) = 0.
\end{equation}
We consider fluctuations of cumulative currents $J^a(t) = \int_0^t \dd t'\, j^a(0, t')$ encoded in the moment generating functions
\begin{equation}
	G^a(\lambda|t) = \langle e^{\lambda J^a(t)} \rangle,
\end{equation}
where the average is taken over a maximum entropy ensemble (flat measure) on ${\rm Gr}_{\mathbb{C}}(k, N)$. By virtue of $SU(N)$ symmetry of the dynamics \eqref{M_eom} and the maximum entropy ensemble, fluctuations of all components are identical, $G^a(\lambda|t) = G^b(\lambda|t)$  $\forall a, b \in \{1, 2, \ldots, N^2-1\}$, which we use to improve sampling statistics.
\subsection*{Isotropic Landau--Lifshitz theory on $\mathcal{S}^2$}
The isotropic Landau--Lifshitz theory, Eq.~\eqref{eqn:LL} of the main text, corresponds to the simplest isotropic Landau--Lifshitz theory on ${\rm Gr}_{\mathbb{C}}(1,2) \simeq \mathbb{CP}^1$ upon identifying the matrix degree of freedom ${\rm M} \in {\rm Gr}_{\mathbb{C}}(1,2)  $ with a unit spin vector ${\bf S} \in \mathcal{S}^2$ as
\begin{equation}
	{\rm M} = {\bf S} \cdot {\pmb \sigma},
\end{equation}
where ${\pmb \sigma} = (\sigma^1, \sigma^2, \sigma^3)^{\rm T}$ is the vector of Pauli matrices.

\end{document}